\documentclass[apj]{emulateapj}

\shorttitle{Outflows in Sodium Excess Objects}
\shortauthors{Park et al.}

\def\nad{Na\,D}
\def\na1{Na\,{\small I}}
\def\mgb{Mg\,\textit{b}}

\def\rchi{$\chi^2_r$}
\def\fnad{\textit{fNaD}}
\begin{document}

\title{Outflows in Sodium Excess Objects}

\author{Jongwon Park$^{1}$, Hyunjin Jeong$^{2,3}$, and Sukyoung K. Yi$^{1}$}
\affil{$^1$Department of Astronomy, Yonsei University, Seoul 120-749, Korea; yi@yonsei.ac.kr\\
$^2$Korea Astronomy and Space Science Institute, Daejeon 305-348, Korea\\
$^3$Korea University of Science and Technology, Daejeon 305-350, Korea}

\begin{abstract}
van Dokkum and Conroy revisited the unexpectedly strong \na1\ lines at 8200\,\AA\ found 
in some giant elliptical galaxies and interpreted it as evidence for unusually bottom-heavy 
initial mass function. Jeong et al. later found a large population of galaxies showing 
equally-extraordinary \nad\ doublet absorption lines at 5900\,\AA\ (\nad\ excess objects: NEOs) and showed that their origins can be different for different types of galaxies. 
While a \nad\ excess seems to be related with the interstellar medium (ISM) in late-type 
galaxies, smooth-looking early-type NEOs show little or no dust extinction and hence no 
compelling sign of ISM contributions. To further test this finding, we measured the doppler 
components in the \nad\ lines. We hypothesized that ISM would have a better 
(albeit not definite) chance of showing a blueshift doppler departure from the bulk of the 
stellar population due to outflow caused by either star formation or AGN activities. 
Many of the late-type NEOs clearly show blueshift in their \nad\ lines, which is consistent with the former 
interpretation that the \nad\ excess found in them is related with star formation-caused gas 
outflow. On the contrary,  smooth-looking early-type NEOs do not show any notable doppler component, which is also consistent with the interpretation of Jeong et al. that the \nad\ excess in 
early-type NEOs is likely not related with ISM activities but is purely stellar in origin.
\end{abstract}


\section{Introduction}
\label{sec:intro}

The behavior of sodium spectral features has garnered much attention as it has become 
known that some galaxies show enhanced \nad\ doublet strengths at 5890 and 
5896\,\AA\  and enhanced \na1\ doublet strengths at 8183 and 8195\,\AA. Numerous 
studies have been performed over the last three decades to understand these lines, but it 
is still unclear exactly how some galaxies exhibit a sodium excess.

Recent research focused on variation of an initial mass function (IMF) has provided an 
interesting possibility. The stellar IMF is usually considered a universal function, but the 
possibility of a non-universal IMF has been raised by several authors 
\citep[see e.g.][]{d08,v08, tetal00,gfs09,tetal10,vc10}. In studies on the Ca\,II triplet at 
8500\,\AA, \citet{setal02} found an anti-correlation between the strength of the Ca\,II 
triplet region and the velocity dispersion for elliptical galaxies and concluded that 
bottom-heavy IMFs are favored \citep[see also][]{cetal03}. Recently, \citet{vc10} reported 
observing a near-infrared \na1\ doublet in the spectra of massive early-type galaxies, 
and claimed that these excesses can be explained by a bottom-heavy IMF 
\citep[see also][]{vc12}.  This implies that massive early-type galaxies should possess 
relatively more low-mass stars.

An alternative solution is also possible. It is well known that  the \nad\ feature is sensitive 
to Na-enhancement ([Na/Fe]). The discovery of non-solar abundance patterns in 
early-type galaxies was first made by \citet{o76} and \citet{p76}. They found extreme 
enhancement of \mgb\ and \nad\ features with respect to calcium and iron peaks, and 
concluded that this was a result of higher metal abundance. \citet{w98} also claimed that 
strong Na features are caused by an overabundance of [Na/Fe] \citep[see also][]{wis11}. 
This is a trivial interpretation and thus not satisfying unless the origin for the enhancement is given clearly.

Furthermore, ISM could also increase the \nad\ line strength. Until the early 1980s it was 
thought that only late-type galaxies had significant ISM. Advances in X-ray and radio 
astronomy, however, have demonstrated that many early-type galaxies also have an 
unignorable amount of ISM. A direct method of measuring hydrogen gas column densities 
to detect ISM is possible via spectroscopy in the ultraviolet, because observations between 70 and 
1000\,\AA\ are sensitive to small amounts of hydrogen and helium in the ISM. If it is not 
easy to obtain such data because of instrumental limitations; an alternative way is to use 
absorption lines such as K\,{\small I}, Ca\,{\small II}, and \na1\ in the more accessible 
optical region of the spectrum. The \nad\ absorption lines, especially, provides a good 
probe of cold ISM in the outflow. According to \citet{cetal10}, \nad\ absorption arises 
from cool gas in the disk and a blueshifted \nad\ absorption is frequently detected in 
star-forming galaxies \citep[see also][]{hetal00}.

To understand the origin of the \nad\ excess, Jeong et al. (2013; J13 hereafter) explored 
the properties of \nad\ excess objects (NEOs) from the seventh data release of the Sloan 
Digital Sky Survey \citep[SDSS;][]{aetal09}, with morphological information obtained 
through visual inspection of galaxy images. 
They found the use of bottom-heavy IMF not
capable of reproducing the observed strength of Na D lines much but instead found it necessary to evoke an {\em ad hoc} enhancement of Na ([Na/Fe$]\,\sim\,0.3$), just as in \citet{w98}. 
Their work does not necessarily rule out the possibility of a bottom-heavy IMF but hints that
there is another missing physics that controls the Na D strengths more importantly \citep[see also][]{yi15}.
Indeed, weak lensing studies (Spiniello et al. 2012; Spiniello, Trager \& Koopman 2015)
suggest that a bottom-heavy IMF (similar to the original Salpeter IMF) helps reproducing the high values of mass derived on lensing galaxies, while the use of Na enhancement was additionally required.

\citet{J13} also found that little dust extinction seems present in smooth-looking {\it early-type} NEOs quoting the OSSY database \citep{ossy11}, which supports the interpretation of stellar 
(rather than ISM) origin for the Na strength excess.
As a further confirmation test, we hereby present the result of doppler measurement on the
shape of \nad\ lines for the J13 samples of NEOs.
If the \nad\ excess is related to ISM, we may detect doppler components in \nad\ line 
shapes due to gaseous outflows in actively star-forming galaxies \citep{cetal10} or active 
galactic nucleus (AGN) galaxies \citep{detal12}.


\section[]{Galaxy Sample}
\label{sec:sample}

The parent sample for this study is the NEOs from J13. The J13 sample is drawn from the 
SDSS DR7 in the redshift range 0.00\,$\leqslant$\,$z$\,$\leqslant$\,0.08 by applying an 
absolute $r$-band magnitude cut-off of $-$20.5 to obtain a volume-limited sample and 
signal-to-noise (S/N) cut-off of 20 to guarantee high quality spectroscopic data. To find 
NEOs, J13 defined a new index, \fnad, which quantifies the \nad\ excess as follows:
\begin{eqnarray}
fNaD = \frac{\rm{Na\,D\,(Observed) - Na\,D\,(Model)}} {\rm{Na\,D\,(Model)}},
\end{eqnarray}
where \nad\,(Observed) is the observed \nad\ line strength and \nad\,(Model) is the 
expected model \nad\ line strength. \fnad\ $\geqslant$ 0.5 is used as a criterion for the 
\nad\ excess, and  0.0 $\leqslant$ \fnad\ $\leqslant$ 0.1 is used to create a control sample. 
The sample galaxies were then morphologically classified via visual inspection. The 
NEOs were carefully assigned to four classes: (1) ordinary early-type galaxies (oETGs), 
(2) peculiar early-type galaxies (pETGs), (3) ordinary late-type galaxies (oLTGs), and 
(4) peculiar late-type galaxies (pLTGs). We note that galaxies with asymmetric features 
and dust patches (or lanes) are classified as peculiar types. In the case of control sample 
galaxies, these are simply divided into two categories: early-type galaxies (cETG) and 
late-type galaxies (cLTG). The details of the sample are described more fully in Section~2 
of J13.

\begin{figure}
\includegraphics[width=0.45\textwidth]{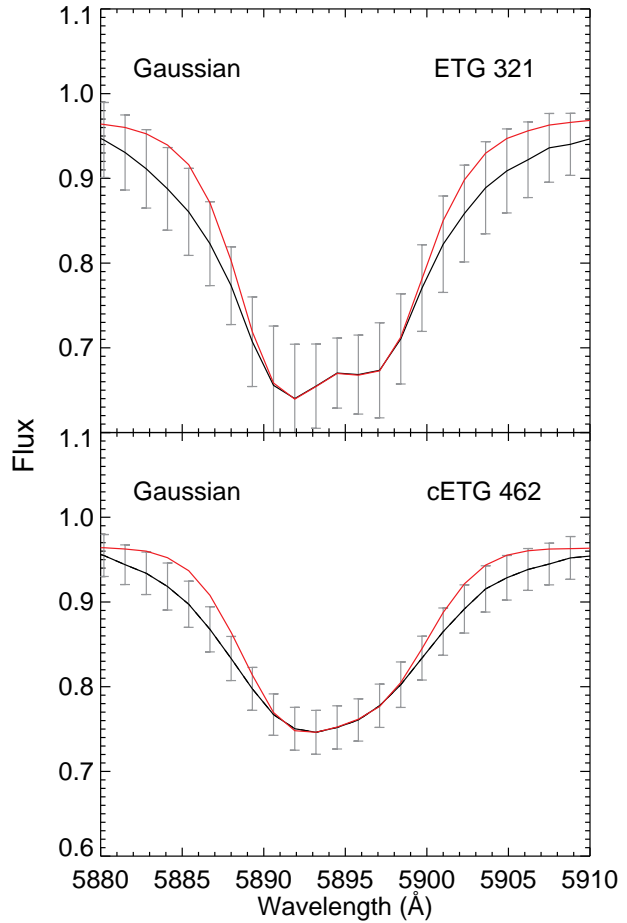}
\caption{The stacked spectra of our early-type NEOs (top) and early-type control sample galaxies (bottom) from the SDSS database (black lines). Only the galaxies with good spectral fit ($\chi_r^2 \leq 3.0$) are shown here. The error bars indicate 1$\sigma$ scatter in the sample. Red lines show the stacked SED of their corresponding fits.}
\label{fig:Gaussian}
\end{figure}

\begin{figure}
\includegraphics[width=0.45\textwidth]{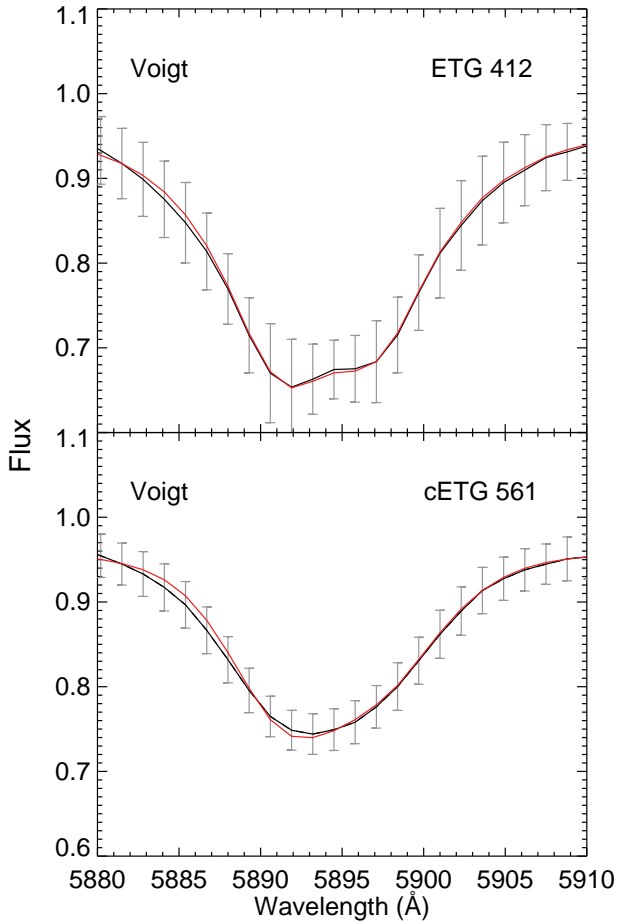}
\caption{Same as Figure~\ref{fig:Gaussian} but based on Voigt fitting. Note that Voigt 
profiles reproduce the shapes of the \nad\ lines much better.}
\label{fig:Voigt}
\end{figure}

\section[]{Fitting Methodology}
\label{sec:method}

As mentioned in Section \ref{sec:intro}, we assume that if \nad\ excess has ISM-origin, \nad\ absorption line would be blueshifted by the effect of outflow such as galactic superwind or AGN outflow. It is known that an \nad\ doublet is a good 
tracer of cold ISM in the outflow. In order to measure the doppler component, we tried both 
Gaussian and Voigt functions to fit each galaxy spectrum near the NaD absorption lines. 
Note that we fit the \nad\ line profile with two Gaussian (or Voigt) profiles. A possible 
criticism is that our fits are mainly mathematical rather than physical. 
However, it is the simplest method that can be used just to investigate whether NEOs 
show a blueshift doppler departure from the bulk of the stellar population or not.

\subsection{Gaussian Fitting}

We first adopted two Gaussian profiles to fit each galaxy spectrum and reproduce the 
shapes of the \nad\ doublet lines using the following formula: 
\begin{eqnarray}
y=a_1 exp{ \left( - \frac { (x-\mu_1)^2 } { 2\sigma^2} \right) } +
a_2 exp{ \left( - \frac { (x-\mu_2)^2 } { 2\sigma^2} \right) },
\label{eq:Gaussian}
\end{eqnarray}
where  $a$ and $\sigma$ represent the line depth and width of the \nad\ absorption 
feature, respectively, and $\mu$ is the position of the centroid. For fitting,  we assume that 
the line widths ($\sigma$) of the two Gaussian profiles are the same because they 
originated from the same galaxy, and two pseudo-continuum band-passes of \nad\ are identified 
as  [5860.625 -- 5875.625\,\AA] and [5922.125 -- 5948.125\,\AA] from \citet{wfg94}.
To measure the doppler components, the quality of our fits to the observed spectra is 
important. Therefore, we calculated a reduced $\chi^2$ (\rchi) near the two dips, and only 
galaxies with \rchi\ $\leqslant 3.0$ are included in our final sample.

Figure \ref{fig:Gaussian} shows the observed stacked spectra (black solid lines) of 
early-type NEOs (ETG) and early-type control sample galaxies (cETG) in the \nad\ region. For comparison, we stacked their Gaussian fits (red solid lines). The models match the observed spectra well near the 
two dips, but there is a marked discrepancy in the fit on the sides (wings). One might be 
tempted to interpret this mismatch as an outflow effect, but if this interpretation is followed, 
then the same result in the early-type control sample is not explainable.

\subsection{Voigt Fitting}

Some lines like Ca II H and K, CaI at 4227\,\AA, \nad, and Mgb show strong 
pressure-broadened wings in the spectra of cool stars. It is known that the Voigt profile is 
particularly well suited to fit the wings of such lines. The Voigt profile is defined by a 
convolution of the Gaussian and Lorentizan functions:
\begin{eqnarray}
V(x)=k\,\tilde{V}(x),
\label{eq:Voigt1}
\end{eqnarray}
\begin{eqnarray}
\tilde{V}(x) = \frac{a \gamma}{\pi} \int_{-\infty}^{\infty} 
\frac{e^{-(x'-\mu)^2/2\sigma^2}}{(x-x')^2+\gamma^2}dx',
\label{eq:Voigt2}
\end{eqnarray}
where $k$ is $a/\tilde{V}_{max}$, $a$ and $\sigma$ denote the depth and width of the 
Gaussian component, $\mu$ is the position of the centroid, and $\gamma$ corresponds 
to the width of the Lorentzian component. The shape of the Voigt profile is highly sensitive 
to the value of $\gamma$, so we defined $k$ as $a/\tilde{V}_{max}$ to restrict the 
$\gamma$ contribution to only the width of the wings. We then adopted two Voigt profiles 
to fit the \nad\ doublet for each galaxy spectrum, using the following form:
\begin{eqnarray}
y=V_1(x)+V_2(x).
\label{eq:Voigt3}
\end{eqnarray}

Observed stacked spectra (black solid lines) of early-type NEOs (ETG) and early-type 
control sample galaxies (cETG) with their stacked fits (red solid lines) obtained using two 
Voigt profiles  are shown in Figure \ref{fig:Voigt}. Figures 1 and 2 show different numbers of galaxies mainly because they show only the galaxies for which $\chi_r^2$ was achieved to be better than 3.0 while this cut depends on the fitting method. In addition, we excluded some fits even if chi squared was small when the fits required the positions of the fitting centroids to be farther than 0.5\AA\ from the pre-assumed positions of the doublets. Some displacement of centroids was allowed in the fitting procedure because of the limited spectral resolution and the uncertainty in redshift determination. Given that the result based on the 
Gaussian fitting shows a marked discrepancy in the fit (see Figure~\ref{fig:Gaussian}), the 
Voigt fitting reproduces the shapes of the \nad\ lines in a markedly better way, especially 
for the wings in the spectra. For a sanity check, the mean value of the width (dispersion) in 
the Gaussian component ($\sigma$ in Equation~\ref{eq:Voigt2}) is 
150\,$\pm$\,50\,km s$^{-1}$, which roughly corresponds to the typical value of velocity 
dispersion of galaxies.

\begin{figure}[t]
\begin{center}
\includegraphics[width=0.45\textwidth]{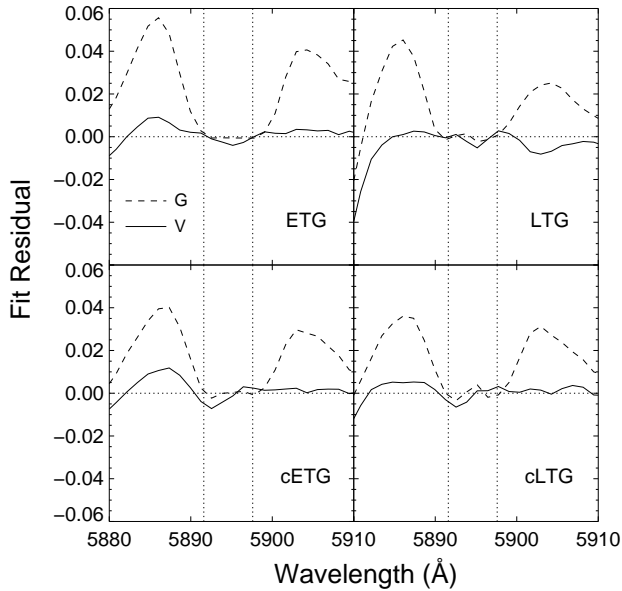}
\caption{Stacked Gaussian (dashed lines) and Voigt (solid lines) fit residuals. Two dip 
positions of \nad\ are shown by vertical dotted lines. }
\label{fig:residual}
\end{center}
\end{figure}

To compare the Gaussian and Voigt fitting methods, we show the fit residuals of our 
sample galaxies according to their morphologies in Figure~\ref{fig:residual}. It should be 
noted that the fit residual based on the Voigt fitting method (solid lines) is below 1\,\%\ 
except in the He\,I region at 5875\,\AA, while the Gaussian fitting method fails to reproduce 
the wings of \nad\ even for the control sample (see cETG and cLTG cases). 

\section[]{Discussion}
\label{sec:discussion}

The broadening of absorption lines is difficult to interpret because the physics of the curve 
of growth is complex. Thus, we focus instead on the blueshift doppler departure from the 
bulk of the stellar population due to outflow caused by either star formation or AGN activity 
by measuring the centroids of the lines. As shown in Section \ref{sec:method}, Voigt profiles provide conspicuously better fits to spectra, so we exploit results of Voigt profile fitting in our final analysis. To measure the doppler component of the \nad\ 
absorption lines, we define as follows:
\begin{eqnarray}
\Delta \mu_1 = \mu_1 - 5891.6\,{\rm \AA},
\end{eqnarray}
where $\mu_1$ and 5891.6  are the left-dip positions of the best-fit for each galaxy and the 
\nad\ absorption lines in vacuum, respectively. If the line shows a blueshift doppler 
departure from the bulk of the stellar population, $\Delta \mu_1$ has a negative value.

\begin{figure}[t]
\begin{center}
\includegraphics[width=0.45\textwidth]{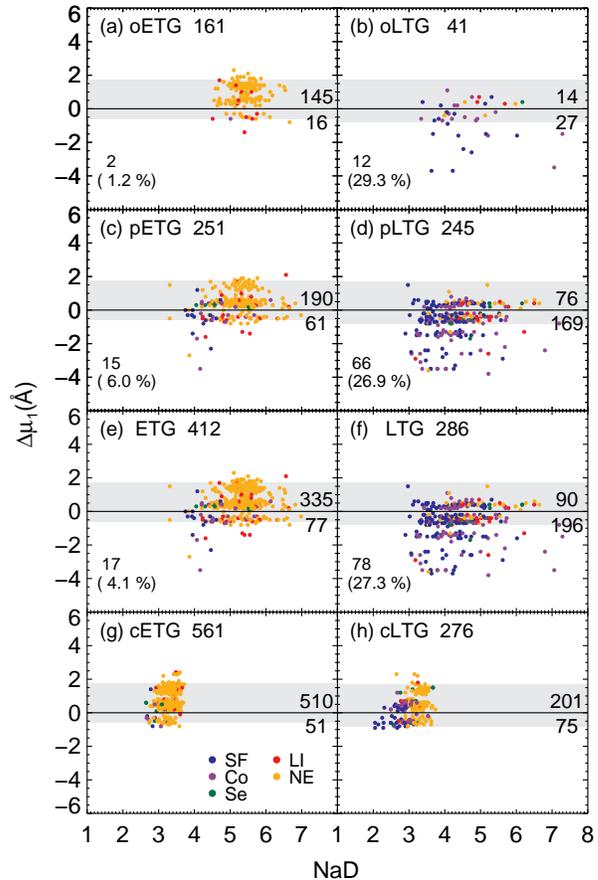}
\caption{Amount of \nad\ absorption line shift with respect to \nad\ line strength. 
In each panel, the number of sample galaxies, the number of galaxies showing blue or redshift are given. Emission-line classification (based on the BPT diagnostic) is also given in different color. The gray band shows the range (encompassing the 95\% of the sample) of doppler shift exhibited by the control sample in panels (g) and (h). For NEOs, the number and fraction of galaxies showing significantly blueshifted \nad\ absorption feature (below the bands) are shown in bottom left corner of each panel (from (a) to (f)). Note that the bulk of our sample galaxies show a systematic red shift by 0.6\,\AA\ compared to the prediction from the vacuum experiment. We only focus on the difference between each sample and the control sample. 
}
\label{fig:mu_vs_NaD}
\end{center}
\end{figure}

Figure~\ref{fig:mu_vs_NaD} presents the amount of \nad\ absorption line shift with respect 
to the \nad\ line strength. To investigate the dominant process that causes this outflow 
component, we use the BPT classification information from J13: star-forming (SF), 
composite (i.e., hosting both star formation and AGN activity, Co), Seyfert (Se), LINER (LI), 
or non-emission line galaxies (NE). Furthermore, the $\Delta \mu_1$ distributions of the 
control sample are shown in gray bands (panels g and h) for easy comparison. The key point in this 
investigation is to check whether the NEOs are so significantly different in the line shift 
compared with the control sample or not.

The left panels of Figure~\ref{fig:mu_vs_NaD} show the early-type cases: ordinary (smooth-looking) 
early-type NEOs (oETG, a), peculiar (non-smooth) early-type NEOs (pETG, c), early-type NEOs 
(oETG\,+\,pETG, e), and early-type control sample (cETG, g). A cursory glance at this 
diagram shows that most early-type NEOs do not seem to show any particular blueshift of 
the \nad\ absorption lines. The fraction showing a shift bluer than those of the control 
counterparts ($\Delta \mu_1$ $\le$ $-$0.6) is roughly 4\,\% (17/412, panel e). This result 
implies that ISM and dust are not likely the main factors causing the increase of the \nad\ 
line strength of early-type NEOs. 

A possible criticism can be made by asking whether the non-shifted \nad\ absorption line 
provides direct evidence of non-ISM. It is known that there should also be a sign of dust 
extinction to allow neutral sodium to survive in the ISM. This implies that, if the \nad\ line 
strength of early-type NEOs is significantly enhanced through ISM effects, these galaxies 
are more likely to show a correlation between dust extinction and \nad\ line strength. 
However, J13 found that there was no correlation between the $E(B\,-\,V)$ values and 
\fnad. On the contrary, the strongest early-type NEOs  were the ones with the lowest dust 
extinction.

It is also worth noting that almost all of the ordinary early-type NEOs (oETGs, panel a) have 
overall ranges of $\Delta \mu_1$ similar to those of the control sample (cETG, panel g). 
For example, the mean values of $\Delta \mu_1$ for oETG and cETG  are 0.87$\pm$0.68 
and 0.78$\pm$0.65, respectively. However, one LINER AGN galaxy below the band is notable. If a galaxy reveals strong emission lines, this galaxy may be a gaseous system. So, there is the 
potential for this galaxy to show the \nad\ excess through ISM effects and to reveal a 
notable doppler component in \nad\ lines due to AGN-powered outflows. The physical 
mechanism by which an AGN could drive molecular gas out of a galaxy is still debated. 
Therefore, a detailed study of this galaxy using integral-field spectroscopy to constrain 
the outflow parameters and ionization mechanisms can shed light on the mechanism of 
gas expulsion from the galaxy and of quenching of star formation in early-type galaxies 
\citep{detal12}.

The right panels of Figure~\ref{fig:mu_vs_NaD} show the late-type 
cases. In contrast to early-type NEOs, 27\,\% of our late-type NEOs (78/286, LTG) clearly 
show blueshift in their \nad\ lines in comparison with their control sample (cLTG).
Furthermore, the overall distribution of late-type NEOs (panel f) is shifted toward more 
negative values than the distribution of the control sample (panel h). For example, the 
mean values of $\Delta \mu_1$ for LTG and cLTG  are $-0.64\pm1.09$ and 0.35$\pm$0.68, 
respectively. This strongly implies that these galaxies have an outflow component. 

\begin{table*}[t]
 \begin{center}
  \caption{A sample of the catalogue of \nad\ excess objects.}
  \begin{tabular}{cccccccccccc}
  \hline \hline
 \multicolumn{1}{c}{SDSS object id} & \fnad\ & \nad \tablenotemark{a} &  Morphology & BPT class \tablenotemark{b} & a1 \tablenotemark{c} & a2 \tablenotemark{c} & $\mu_1$ \tablenotemark{d} & $\mu_2$ \tablenotemark{e} & $\sigma$ \tablenotemark{f} & $\gamma$ \tablenotemark{g} & \rchi\ \\
& & (\AA) & & & & & (\AA) & (\AA) & (\AA) &  \\
(1) & (7) & (8) & (15) & (16) & (17) & (18) & (19) & (20) & (21) & (22) & (23)  \\
 \hline
587727179531354122 & 0.53 & 4.65 & oETG & Quiescent & 0.32 & 0.24 & 5892.8 & 5898.5 & 1.0 & 2.3 & 0.9 \\
588009365862285317 & 0.53 & 4.76 & pETG & LINER & 0.20 & 0.19 & 5891.1 & 5897.3 & 2.7 & 3.0 & 0.8 \\
587727179528339473 & 0.74 & 4.96 & oLTG & LINER & 0.31 & 0.28 & 5892.3 & 5898.3 & 1.5 & 2.3 & 1.9 \\
587737827826204741 & 0.94 & 4.04 & pLTG & Star-forming & 0.21 & 0.16 & 5889.0 & 5894.9 & 2.3 & 3.3 & 1.9 \\
 \hline
 \end{tabular}
 \label{tab:all_sample}
 \end{center}
 \bf{~~Notes.}
\tablenotetext{a}{Observed line strength.}
\tablenotetext{b}{Emission line classification.} 
\tablenotetext{c}{Depth of Gaussian component.}
\tablenotetext{d}{Centroid of the left Gaussian component.} 
\tablenotetext{e}{Centroid of the right Gaussian component.}
\tablenotetext{f}{Standard deviation of Gaussian component.} 
\tablenotetext{g}{Half the FWHM of Lorentzian component.} 

\end{table*}

We checked the BPT classification for the 78 late-type NEOs with blueshifted 
\nad\ lines. Of 78,  41 and 30 galaxies are star-forming (SF) and composite 
(Co) galaxies, respectively, and only 5 galaxies are AGNs. Such findings suggest that 
the \nad\ excess found in these galaxies is related with star formation-caused gaseous 
outflows (galactic winds), which play an important role in the evolution of galaxies by 
removing/heating of cold gas in galaxies.

We thus conclude that early-type NEOs and late-type NEOs have completely different 
mechanisms underlying their \nad\ excess. Many late-type NEOs clearly show blueshift in 
their \nad\ lines, which means that their \nad\ excess is related with ISM. On the other hand, 
early-type NEOs do not show any significant doppler component.
While this does not necessarily rule out a possibility of ISM origin, it would be much more
natural to conclude that their excessive \nad\ is of stellar origin.
To facilitate follow-up observations of these exciting objects, we provide a catalog of the 
sample galaxies presented in this paper in Table~\ref{tab:all_sample}.
 
\section*{Acknowledgments}

HJ acknowledges support  from the Basic Science Research Program through the National 
Research Foundation of Korea (NRF), funded by the Ministry of Education 
(NRF-2013R1A6A3A04064993). SKY acknowledges support from the National Research Foundation of Korea(Doyak 2014003730). This study was performed under the DRC collaboration between Yonsei University and the Korea Astronomy and Space Science Institute.


\clearpage

\end{document}